# Digital Twins for Ports: Derived from Smart City and Supply Chain Twinning Experience

**ROBERT KLAR[1,2], ANNA FREDRIKSSON[1], and VANGELIS ANGELAKIS[1]**
[1]Department of Science and Technology, Linköping University, Campus Norrköping, 60 174, Sweden
[2]Swedish National Road and Transport Research Institute (VTI), SE-581 95 Linköping, Sweden

Corresponding author: Robert Klar (e-mail: robert.klar@liu.se).

This work has been supported by Trafikverket Sweden as part of the Triple F (MODIG-TEK) project

**ABSTRACT** Ports are striving for innovative technological solutions to cope with the ever-increasing growth of transport, while at the same time improving their environmental footprint. An emerging technology that has the potential to substantially increase the efficiency of the multifaceted and interconnected port processes is the digital twin. Although digital twins have been successfully integrated in many industries, there is still a lack of cross-domain understanding of what constitutes a digital twin. Furthermore, the implementation of the digital twin in complex systems such as the port is still in its infancy. This paper attempts to fill this research gap by conducting an extensive cross-domain literature review of what constitutes a digital twin, keeping in mind the extent to which the respective findings can be applied to the port. It turns out that the digital twin of the port is most comparable to complex systems such as smart cities and supply chains, both in terms of its functional relevance as well as in terms of its requirements and characteristics. The conducted literature review, considering the different port processes and port characteristics, results in the identification of three core requirements of a digital port twin, which are described in detail. These include situational awareness, comprehensive data analytics capabilities for intelligent decision making, and the provision of an interface to promote multi-stakeholder governance and collaboration. Finally, specific operational scenarios are proposed on how the port's digital twin can contribute to energy savings by improving the use of port resources, facilities and operations.

## I. INTRODUCTION

**P**ORTS play a central role as part of the global transport chains, bridging the sea and land transport legs. As part of diverse transport and logistics networks, ports ought to support efficient transshipment and intermodal transfer [1]. The importance of their role in the global transport system is increasing, as 90% of world trade passes ports on its way from origin to destination [2]. Ports are also at the center of more local economies, since being gateways to global trade they support domestic networks and can thus be considered as a catalyst for the economic development of regions and entire countries [3]. With the steady growth of maritime transport and global sustainability efforts, ports are under increasing pressure to improve their profitability, environmental friendliness, energy performance and efficiency. [4]. The dynamic and competitive character of the maritime transport chains drives the application of new technologies and innovation to enhance performance and increase co-operation and transparency, and attract new business [5]. Ports are economic catalysts for neighboring cities that can facilitate market integration and agglomeration of services and generate social and economic benefits. Likewise, cities provide essential resources and infrastructure to ports, and ports provide goods, raw materials, and other transportation services for urban development [6]. Primarily, ports serve as transshipment hubs in transportation chains by linking different modes of transportation, acting as facilitators of international trade as transportation demand varies spatially, and facilitating temporary storage to match size differences between different modes of transport [7]. The connection between port and hinterland is an area with room for improved efficiency in operations and sustainability. Since the majority of freight transport between port and hinterland is carried out by truck, the presence of a port can lead to congestion on the (peri-)urban road network caused by hinterland traffic to and from the port area. The resulting congestion on the urban road network not only affects the quality of life among citizens in vicinity, but also the competitiveness of ports [8]. Ports are therefore closely linked to (smart) cities as well as to supply chains, as they are both nodes in transport chains [9] and providers and users of urban infrastructure, as ports increase the global connectivity of cities while using their infrastructure for hinterland transport [10]. Apart from their functional linkage, all three are also characterized by a set of commonalities, mainly their complex interplay of various interconnected processes and actors.

Preparing for the future, many European ports, are focusing on safety, efficiency and sustainability and have been launching efforts to provide a complete and up-to-date overview of port activities via digital twinning [11]. Twinning in this context aims to enhance ports real-time situational awareness for static, moving, human-controlled or autonomous entities and artifacts, by bringing together geographic, process, or sensor information, be it historic or real-time.

Taking a step back from ports, Digital twins (DT) enable simulation of systems' behavior and have been referred to as a "quantum leap in discovering and understanding emergent behavior" [12]. Originally developed to support manufacturing, digital twinning has attracted a great deal of attention, especially from the perspective of systems' engineering [13]. The main focus of several industries such as transportation, security, aerospace, manufacturing, and many others, is to achieve optimal performance, reliability, robustness, and efficiency among various of their systems with different characteristics, to accomplish a common objective [14]. This has led to a growing interest in integrating different independent systems to enhance their overall capabilities and performance. Within this context, the idea of the digital twin is to be able to design, test, manufacture, and eventually use the virtual version of complex systems [12]. Such systems relevant to ports encompass port congestion systems [15], information systems [16], and port traffic flow simulation systems [17]. Examples from the smart city domain, which is closely functionally connected to the port, include monitoring systems [18], intelligent transportation systems [19], urban fine management systems [20] and system of systems approaches [21], [22].

Although the concept of the DT has largely evolved since its coining in 2002 [12] and initial successful digital twinning implementations within the port domains exists, there is still a lack of standardization, methodologies, and tools for the development and implementation of DTs [23]. Furthermore, the concept and content of DTs lack a precise, uniform definition and description [24]. This results in problems with a uniform implementation of the DTs in ports. An additional obstacle is that different port actors usually keep a wide array of, practically, vertical information systems (i.e. with limited or no actual interconnection between them), due to the large number of actors in the port processes. A further obstacle is the large number of port actors with a variety of information systems with limited or no interconnection. This results in a lack of data exchange, leading to individual operators inability to plan required capacities (short and long term), making it difficult to accurately predict when a port a call will occur and hence what resources are needed when [25]. Furthermore, although all ports share common characteristics, such as providing services to freight and ships, they often differ wildly in terms of size, geographic characteristics, governance, functionality, and specialization [1]. Therefore, there is no one-size-fits-all solution for a port DT, but nevertheless a set of requirements and DT aspects relevant for most ports regardless of their specificities should be able to be identified, which will be the further discussed in this paper.

To summarize, in contrast to traditional DT applications, ports are differentiated by their complex interplay of independent actors, processes, and activities that are because of present lack of connection between systems subject to uncertainty [26]. However, these challenges are also shared with smart cities and supply chains. In addition, the port as a part of the transport chain in supply chains and as physically interconnected with city infrastructure is functionally intertwined with both of these areas. Although research papers investigating the potential of twinning of port elements have contributed significantly to understanding the potential of digital twinning in ports, a holistic digital twinning approach that covers the entire functionality of the port in terms of its fundamental role as a node in the transport chain of global supply chains and as an important transport infrastructure in the smart city it is still in its infancy.

Based on commonalities of similar DT requirements as well as functional linkage with each other, this paper intends to discuss the requirements and structure of a port DT, building on the knowledge and developments of DTs in the urban and supply chain contexts. More specifically, this paper contributes to the DT research by the following:

- A thorough analysis of DT definitions across two application domains, that of smart cities and of the supply chain, yielding a discussion of core characteristics, enablers and potential usage in the port domain
- A characterization of a port DT through a comprehensive analysis of its processes and characteristics, leading to the identification of three core requirements (taking into account the interrelationships and commonalities with smart city and supply chain DTs)
- Proposing operational strategies on how a port DT can contribute to energy savings through the optimal use of port facilities and equipment and intelligent linking of port processes
- Discussing potential barriers delaying large-scale digital twinning of the port and possible consequences of the same

The paper is organized as follows. In section II we outline the methodology of the paper. Section III addresses how definitions of the DT have developed, and differ depending on their application domain. Building on the characteristics of the different DT definitions discussed in Table 2, Table 3 presents the cross-domain characteristics. In section IV, digital twinning implementations of the smart city and the supply chain and the decisive role of the port within these application domains are illustrated in Table 4. DT implementations for the smart city and the supply chain are thus discussed in more detail in subsections IV-A and IV-B resulting in Tables 5 and 6. Based on an intensive study of all operational processes in the port considering the DT characteristics discussed in section III, and with respect to commonalities with smart city or supply chain aspects, three main objectives of the port's DT are defined and discussed in depth in chapter V. These three main objectives compromise the importance of situational awareness in subsection V-A1, data-driven decision-making in subsection V-A2, and the importance of providing tools for enhanced collaborative actions in subsection V-A3. The insights gained in section V will be further elaborated to gain an understanding of how a port DT can contribute to energy savings resulting in chapter V-B. Concrete practices are presented in Table 7 on how the DT can contribute to the effectiveness of port operations by optimizing the use of port facilities and equipment and aligning processes. Section VI further discusses potential barriers in implementing a port DT. The core results are subsequently discussed and summarized in section VII and IX respectively.

## II. METHODOLOGY

The purpose of this section is to describe the underlying methodology and the research process of this paper. The paper is a conceptual paper identifying interpretable patterns and relationships withing existing digital twin solutions. According to Hulland in [27], conceptual papers primary focus on either (1) synthesizing existing knowledge or (2) developing new ideas. Thereby, conceptual papers do not include data, as their focus is on integrating and revealing new relationships between constructs [28]. The conceptual development in this paper is based on a descriptive literature review [29].

Drawing on the work by Jaakkola in [30] which presents four types of research design in conceptual papers, this paper combines a theory synthesis and a conceptual model approach. The former aims to provide a new or expanded view of a concept or phenomenon by linking previously unconnected or incompatible pieces together in novel ways, thus summarizing and integrating existing knowledge about a concept [30]. According to MacInnis in [31], summarizing helps researchers see the forest for the trees, while integration allows researchers to see a concept or phenomenon in a new way by transforming previous findings and theories into a new, higher-level perspective that connects phenomena previously considered distinct. This paper first analyses the core DT characteristics and second it discusses how ports can learn from experiences in the smart city and supply chain domains. Third, a conceptual model, aiming to provide a theoretical framework predicting the relationships between concepts [30] is elaborated. The model aims to provide a roadmap to previously unexplored relationships between constructs, and introduce new constructs [30], [31], by outlining cross-domain DT characteristics that lead to the identification of port-related DT systems.

TABLE 1: Applied search queries in Web of Science

| Index | Search query | Records |
| --- | --- | --- |
| 1 | (port OR harbour OR terminal) AND ("digital twin*") | 132 |
| 2 | ("digital twin*") AND (survey OR review OR applications) | 3359 |
| 3 | (city OR urban) AND ("digital twin*") | 760 |
| 4 | (("supply chain") AND ("digital twin*")) OR (digital AND ("supply chain") AND (twin*)) | 256 |

The following paragraph presents the research approach. An extensive cross-domain literature review was conducted to analyze the current state of the art of DTs. The search strategy used was a combination of search queries in Web of Science and Google Scholar to find a wide selection of papers and snowballing to identify relevant papers not making use of the keywords used in the search queries. The aim was to get an extensive overview of a scattered area. Table 1 illustrates the applied search queries and their respective number of records in Web of Science. In addition, the search queries were further used in Google Scholar. A further sub-selection of papers was selected based on the relevance of the scanned titles and abstracts, as well as further discussion between authors. In total, 82 DT-related papers were included in this paper.

The papers were selected with the aim to identify definitions and goals of DTs across domains. The commonalities cross domains were identified as core characteristics, which were further explored to extract commonalities and differences between domains to identify solutions transferable to the port context. During this process it was identified that the port is most comparable to supply chains and smart cities, due to its large number of complex inter-liked processes and actors, and therefore high level of uncertainty. Thus, as a next step an in-depth investigation of DT implementations in the smart city and supply chain domains was conducted. Thereafter to identify functionality and requirements of a port DT supporting port efficiency as a whole, a detailed examination of the numerous, interconnected port processes and the complex interplay of different actors in the port were analyzed. Finally, to exemplify opportunities of resource efficiency, examples of how DTs can save energy by intelligently controlling and linking port processes were elaborated. The research process, which is mirrored in the structure of the paper is presented in Figure 1.

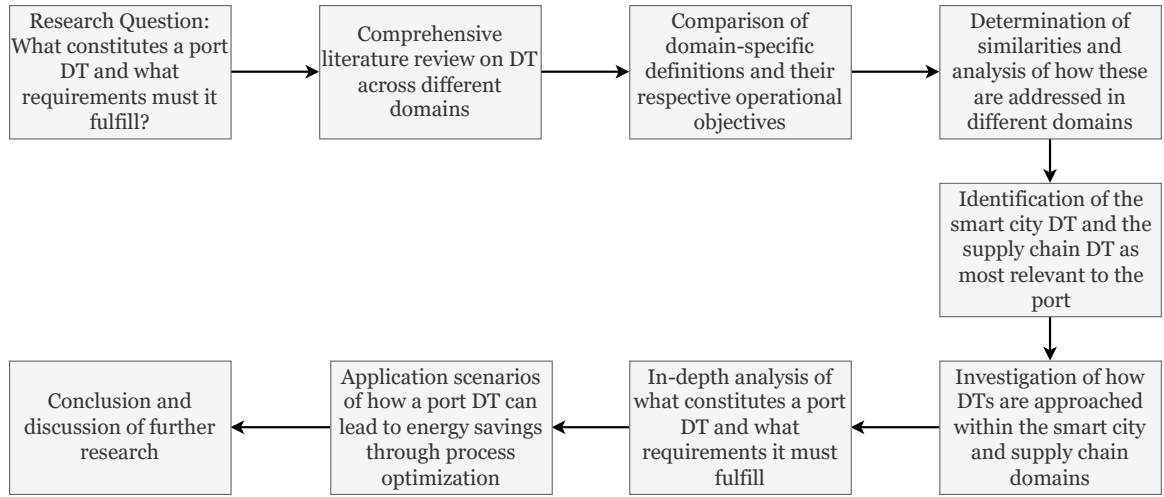

FIGURE 1: The research process illustrates the steps taken for the implementation of the paper.

## III. LITERATURE REVIEW OF DIGITAL TWINS

According to a recent report by the European Institute of Innovation and Technology (EIT) [32], the engineering industry now sees the dire need for DTs. This is evident by projecting the global DT market to reach a value of USD $ 35.8bn by 2025, a tenfold increase in value compared to 2019. Due to the constant evolution of the technologies that enable the DT and the ongoing development itself, as well as varying definitions and interpretations of what constitutes a DT, there is still confusion about what should constitute a DT. Therefore, the purpose of this section is to provide a historical overview of the development of the DT (Figure 2), compare domain-specific definitions of the DT (Table 2), present the most important functions and features and demonstrate how these are addressed by different domains (Table 3).

### A. HISTORY OF DIGITAL TWINS

The concept of "twinning" dates back to the Apollo program of the National Aeronautics and Space Administration (NASA), where at least two identical space vehicles (twins) were built to mirror the real operating conditions for simulation of the real time behavior of the space vehicle during the mission [33]. The origin of DTs can be traced back to a University of Michigan presentation to industry in 2002 that suggested the creation of a Product Lifecycle Management (PLM) center [12]. Although the concept was originally formulated as a "Conceptual Ideal for PLM", it already contained all the elements of the DT: real space, virtual space, the link for data flow from real space to virtual space, the link for information flow from virtual space to real space, and virtual subspaces [12]. The development of the DT was accompanied by the goal of providing "three of the most powerful tools in the human knowledge toolbox" [34]: (1) Enhanced **conceptualization** by providing enhanced situational awareness, (2) extensive possibilities for **comparison** to benchmark different product prototypes based on a set of ideal characteristics, and (3) improved **collaboration** by viewing any physical product at any stage. Although the concept was well received, it took several years for the concept to fully enter academia and industry, as the enabling technologies to create the DT were still in their infancy. However, these technologies, especially advances in IoT, experienced exponential growth since then. In 2012, the concept of DTs was revised by NASA, which used DTs for spacecraft operations and maintenance. They defined the DT as "an integrated multiphysics, multiscale, probabilistic simulation of a complex system that uses historical data, real-time sensor data, and physical models to mirror the life of its corresponding twin" [35]. The concept of the DT then heavily impacted the manufacturing domain as a tool to provide guidance in how products are designed, realized, used, and disposed [36]. The attention of both academia and professionals towards DTs even lead to a call for a new smart manufacturing era and many strategies have been discussed on how to achieve smart manufacturing [37]. The growing interest in the DT is also reflected in the fact that DTs were ranked by Gartner as one of the top 10 technology trends with strategic value between 2017 and 2019 [38]–[40]. The increasing popularity of the DT also led to applications going far beyond manufacturing. In 2017, the Data for the public good report by the UK national infrastructure commission outlined a roadmap towards a national DT to monitor UKs infrastructure in real-time, and to simulate the impacts of possible events such as natural disasters or an extension of the train network [41]. Since this report, there have been large efforts to transfer the principle of the DT has been towards cities [42] or entire supply chains [43], [44]. A brief timeline of the key events leading towards DTs for ports, including the integration of existing DT approaches from the ports of Rotterdam [45] and Hamburg [46], is illustrated in Figure 2.

Presently, DTs are popping up everywhere [47], in smart manufacturing, building management, smart city, construction, engineering, transport and medicine, to name just a few prominent examples. In these domains, DTs are used to better understand, control, and optimize the behavior of complex systems, either at design time (e.g., to explore the design space) or at runtime (e.g., to improve performance/productivity or avoid failures) [48].

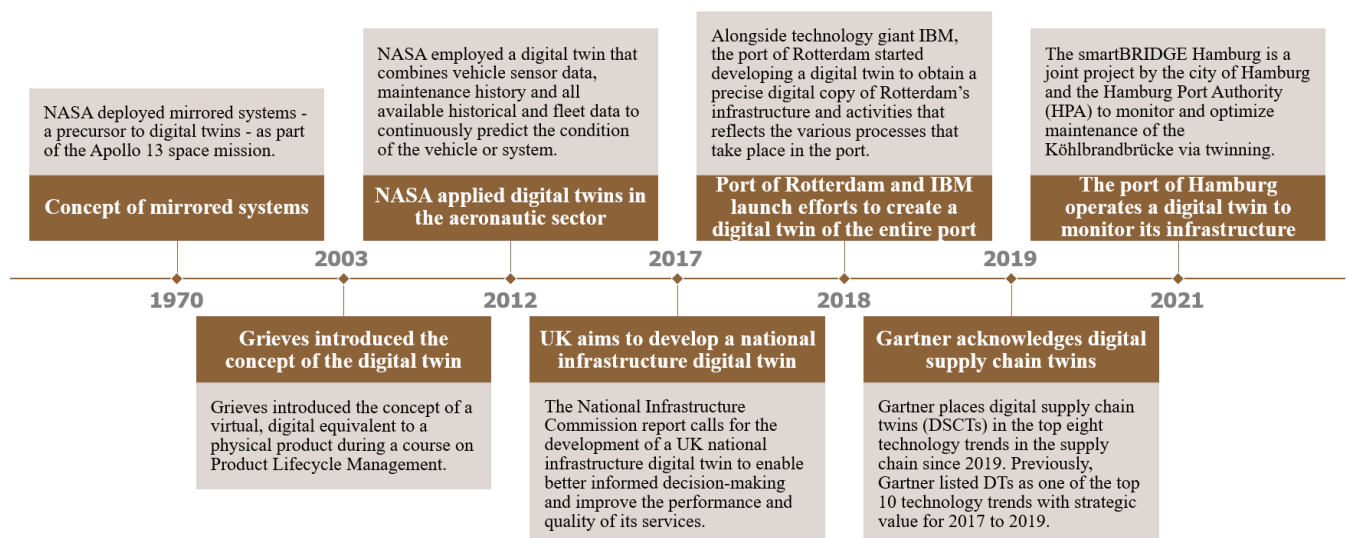

FIGURE 2: Timeline of DT highlights leading to a holistic port DT.

### B. DEFINITION AND CHARACTERISTICS OF DIGITAL TWINS

Twinning of physical assets, or processes represents a step in the process of digitization and, as such, has been evolving together with technologies supporting its realization (e.g., sensor technology, Internet of things (IoT), cloud computing, big data analytics, and artificial intelligence (AI)) in the last twenty years [49]. DTs are considered to be the pillar of Industry 4.0 and the innovation backbone of the future as they bridge the virtual cyberspace with physical entities [50]. Several definitions of DTs (DT) consider it a virtual representation (replica) of an actual system (AS), which can continuously update with real-time data throughout its lifecycle and can interact with and influence the AS [48]. However, since DT is the subject of study in several disciplines and a tool applied across different disciplines, where practitioners understand it from their unique professional perspective, there is no uniform definition. Comparing different definitions, it is evident that the more recent definitions of DT focus on dynamics, learning, and evolution, rather than just being digital shadows of static objects in the real world [21]. Following the findings of Evans et al. in [32] and Botín-Sanabria et al. in [51], it is appropriate to assess DT solutions according to their maturity. In this context, six successive maturity levels are proposed, which can be summarized as Reality capture (level 0), Replication (level 1), Connection (level 2), Synchronization (level 3), Interaction (level 4) and Autonomous (level 5). Based on the comprehensive evaluation of existing DT solutions by Botín-Sanabria et al. [51], most DT concepts are still at initial stages, and few have started integrating real-time data streams, because capturing, filtering, and processing data in real time is a major challenge, and device malfunctions and poor calibration can lead to anomalies or missing data points. An updated maturity assessment, applied to evaluate three innovation-leading ports is presented in [52]. In their maturer versions, DTs are more than just Building Information Modeling (BIM) or 3D models. They can then serve as a data resource that enhances the design of new facilities and the understanding of the condition of existing facilities, verify the as-built condition, perform "what-if" simulations and scenarios, or provide a digital snapshot for future works [32]. Consequently, a fully developed DT is expected to have elements of self-adaptiveness in combination with machine learning, simulation, and data processing to enable accurate prediction of specific properties related to performance [48]. Drawing on the work of Jiang et al. [50], some of the core aspects of the DT are summarised here.

- **Components:** A DT and its physical counterpart consists of physical entities, virtual models, physical-digital connections, data, and services .
- **Temporal span**
  - **Complete lifecycle:** A DT is designed to synchronize its physical counterpart at all stages covering design, prototyping, manufacturing, deployment, maintenance, and disposal.
  - **Changing requirements:** Depending on the application domain, the DT may have different requirements depending on the time cycle. In manufacturing, iterative optimization and data integrity are crucial in the design phase, while real-time monitoring, process evaluation and optimization are essential in the production phase, while predictive maintenance and fault detection and diagnosis are important in the service phase [53].

- **Increased value over time:** A DT is a self-improving system that can be progressively improved and extended through the increasing accumulation of data and knowledge over time.
- **Functional scope**
  - **Modelling:** A DT is a grouping of models and algorithmic components jointly describing a complex system. The virtual representation is constantly updated by data streams from its physical counterpart or by results of simulations. The comprehensive modeling involves all aspects of the physical entities as well as the prediction of additional likely outcomes to test, for example, what-if scenarios and for predictive maintenance.
  - **Visualization:** A DT enables a digital replica of all static and dynamic processes as well as the components of its physical counterpart. With its distinctive visualisation capabilities, the DT allows permanent monitoring of the processes and thus contributes to situational awareness and enables the cooperation of different actors in the digital space.
  - **Interaction:** A DT is characterised by its Bi-directional character. Compared to a digital shadow that simply monitors the actual system in real-time, the DT directly influences the actual system based on its actions, changes and predictions. Furthermore, the DT interacts with other fleet members if there are multiple instances, as well as with human operators.
  - **Synchronization:** The DT is continuously updated in a timely manner by various components and processes of the actual system whenever needed. With sufficient communication bandwidth, this ensures a consistent up-to-date virtual representation and is required for a number of ongoing online tasks such as scheduling, control and optimisation.

Although there are already several existing papers comparing differing DT definitions, such as Barricelli et al. in [81], Wu et al. in [82] or Liu et al. in [83], these are either unstructured [83], structured by time of publication [53], [84], [85], by application [86] or by DT key points [81]. However, to our best of knowledge, there is no paper (1) comparing DT definitions including both smart cities and supply chains as application domains, and (2) outlining their purposes as performed in the case of this paper. The resulting comparison of different domain-based definitions and purposes of DTs in Table 2) reveals that different application domains have different requirements, particularly in terms of the scale, frequency of updates and predictive capabilities. This distinction is underlined by Mylonas et al. [21] who point out that scale is one of the fundamental differences between DTs in smart manufacturing and those in smart cities, as smart cities are essentially systems of systems and the complexity and heterogeneity of DTs at the urban scale may be orders of magnitude greater than their industrial counterparts. Another example of a highly complex system is the supply chain, which consists of an extensive network of actors and processes. However, compared to a manufacturing or healthcare application, at least some of the required information (e.g., the position of the ship or the status of a warehouse) is needed in a timely manner, when updates are required, rather than a real-time fashion, whereas a DT monitoring a patient's health parameters or performing predictive maintenance analysis of a running high-performance product requires information about all required parameters in real time. Consequently, a domain-based definition of DTs might be more advantageous than a general definition as this cannot apply to all domains. Consequently, there is a need to compare how different DT solutions across domains approach the DT key objectives identified in Table 2. The resulting DT core objectives valid across domain can be summarized as (1) threat detection, (2) energy savings, (3) cost reductions, (4) increased performance, and (5) enhanced collaboration. Table 3 addresses this need by comparing how the five selected domains presented in Table 2 address the core DT objectives identified above to obtain an understanding of which domain approaches might be of interest for adaptation to the port.

Comparing the different domain-based definitions in Table 2 and the different solution-oriented approaches to the identified core objectives in Table 3, it becomes apparent that the port DT has similar characteristics (such as complex interplay of actors and processes) and requirements as the smart city and supply chain twins, and could learn most from their solution approaches. Similar to the supply chain, the port also needs to accurately estimate the arrival time of incoming ships and trucks, promote smooth cooperation between the different actors, monitor and detect disruptions in real time and define measures for the time of disruption and recovery. In addition, similar to the smart city, the port has the requirement to evaluate the risk of flooding, ensure smooth traffic flow while avoiding congestion, establish the best operating scheme through simulation, and achieve energy savings while minimising failures through optimal use of facilities and resources. By identifying the smart city and supply chain research approaches as most relevant, these are discussed in more detail in the following two sections.

## IV. THE PORT AS AN INTERSECTION OF THE SMART CITY AND THE GLOBAL SUPPLY CHAIN

This section clarifies the interrelation of the port to the smart city and the supply chain by illustrating their interplay in Figure 3. Subsequently, existing research approaches and twinning solutions in the domain of smart cities IV-A and supply chains IV-B will be explored in depth, as these are particularly relevant to the port. By taking a deep dive into these two domains, insights for the port DT can be drawn from examining digital twinning solutions and approaches in these two related domains.

In academia as in practice, seaports are viewed as nodes in transport chains [9] as well as parts of the smart city

TABLE 2: Domain based definitions of DTs

| Domain | Definition | Purpose | Key characteristics | References |
|---|---|---|---|---|
| Manufacturing | "A digital twin is an integrated multi-physics and multi-scale simulation of a product/system that can model the mechanical, electrical, software, and other discipline-specific properties across its lifecycle" [54]. "Through high-fidelity modeling, real-time interaction and data fusion, DT can reproduce a physical asset or process accurately in the digital world and enable more effective monitoring, optimization, and prediction of the physical counterpart throughout its lifecycle" [55]. | "The most direct value of implementing digital twins is to replace the costly pure-physical commissioning and test ... The established digital twin in the SMS design phase can also be further utilized for monitoring, optimizing activities, diagnostics, and prognostics in the following-up lifecycle phases of SMS" [54]. "The advantages of DT can empower researchers and practitioners to develop new methods, tools, and technologies in design, production, and service, eventually leading to more innovations" [55]. | Product optimization, Threat detection, Monitoring, Diagnostics and prognostics | [21], [54], [55] |
| Smart City | "An urban digital twin can be best characterized as a container for models, data, and simulations" [42]. "The city digital twin is anticipated to construct a link with the real city or the physical counterpart to enhance the visibility of the city and the understanding and analysis of the city's events and operations. For that purpose, the city digital twin is perceived as enabling technology to promote situational awareness for city management and to provide a city information model; that is, the city digital twin can collect, monitor, and manage city data" [56]. | "Using an urban digital twin in virtual reality is not only a novel way of using smart technologies for collaborative planning processes, but also facilitates consensus-building among participants with different backgrounds" [42]. "Use-case scenarios like projecting in real time the effect of changes in traffic flows inside urban areas on the mobility of citizens, or calculating the effect of a rainstorm on the level of a river's water and creating potential risk management scenarios, are beginning to sound as a plausible DT application in smart cities" [21]. | Collaboration, Decision-making, Accurate mapping, Virtual-real interaction, Intelligent feedback | [21], [42], [56], [57] |
| Supply Chain | "A digital logistics twin or digital supply chain twin (DSCT) is a digital dynamic simulation model of a real-world logistics system, which features a long-term, bidirectional and timely data-link to that system. The logistics system in question may take the form of a whole value network or a subsystem thereof. Through observing the digital model, it is possible to acquire information about the real logistics system to draw conclusions, make decisions and carry out actions in the real world" [58]. "A digital SC twin is a model that represents the network state for any given moment in time and allows for complete end-to-end SC visibility to improve resilience and test contingency plans" [59]. | "A DSCT acts as a tool for decision-makers in logistics and supply chain management to holistically improve logistics performance along the whole customer order process through data-driven decision-making" [60]. "A digital twin gives analysts the possibility of experimenting with the SC's computer prototype to test what-if scenarios and quantify the effects of changes" [59]. "The core functionality of a DTSC is to provide an integrated and holistic view, which enables all stakeholders to contribute jointly to the creation across different supply chain stages" [61]. | Interaction, Simulation, Real-time planning, Visualization, Collaboration | [58]–[61] |
| Built Environment | "An integrated software solution to manage static and dynamic information of a built asset across its lifecycle phases. It usually provides a realistic digital representation of the physical asset, generated by enriching the geometric or graphical data with support from building automation systems (BAS), sensors, internet of things (IoT) components, and other feedback systems informing about the asset, its occupants, or its environment" [62]. | "DT technology enables practitioners and managers to improve resilience, manage risk, and save energy and resources" [63]. "As a valuable digital asset of the built environment, digital twins have the potential to help us in a variety of cases, including facilities management and operation, asset condition monitoring, sustainable development, etc. Especially with decision making, digital twins can provide all stakeholders of the built environment more reliable and useful information" [64]. | Facility management, Enhanced safety, Energy savings, Monitoring, Collaboration | [62]–[64] |
| Healthcare | "A digital twin is a digital replica of real world devices, processes or even persons. The technology draws on domains like machine learning, artificial intelligence and software analytics to provide a dynamic digital representation of its physical counterpart. Thereby, it uses data provided for example by Internet of Things (IoT) sensors as well as information coming from past machine usage and human domain experts" [65]. | "Digital Twin studies in the field of health have mainly focused on the creation of twins of the organs in the human body, understanding the cell behavior in the body, and applying appropriate medicines and treatments" [66]. | Prevention, Treatment, Understanding of disease, Cost reduction | [65], [66] |

TABLE 3: Commonalities and differences of the domain-based approaches

| DT objective | Domain | How is it approached |
|---|---|---|
| Threat detection | Manufacturing | Smart manufacturing in conjunction with a digital twin has the capabilities of intelligent sensing and simulation, allowing to monitor the status of products and production equipment in real time and predict potential failures in a timely manner [67]. |
| | Built Environment | Application of a Bayesian change point detection methodology that processes the contextual characteristics of behavioral data to identify asset anomalies by cross-referencing them with external operational information [68]. |
| | Smart City | The digital twin can help with epidemic prevention and control based on its disease transmission model, using spatiotemporal proximity, AI analysis, and other technologies to determine the occurrence of the epidemic and identify those in close contact [57]. |
| | | Urban Digital Twins are also used to simulate flooding scenarios by predicting how a rise in urban waters would spill over onto surrounding streets. This flooding information can be used for effective urban evacuation and for the the placement flooding countermeasures [69]. |
| | | Digital twin simulations that rely on data collected from sensors distributed on roadways can aid regulating the traffic flow to reduce the risk of road congestion and traffic accidents. They also have potential to assess the safety aspects of autonomous driving. [70]. |
| | Supply Chain | The digital twin creates a virtual asset based on BIM and simulates various logistics scenarios based on a GIS-based routing application to predict various risks that may occur in the logistics process and calculate an estimated time of arrival [71]. |
| | | The combination of model-based and data-driven approaches of the digital supply chain twin enables predictive and reactive decision making to monitor and detect disruptions in real time and determine actions for the time of disruption and recovery [59]. |
| Energy savings | Manufacturing | Based on empirical models, an improved energy consumption model is proposed that effectively reflects the relationship between machining parameters and energy consumption in machining processes and contributes to sustainable manufacturing [67]. |
| | Built Environment | The digital twin model for evaluating energy-saving strategies is based on the integration of the building's physical environment and a database containing the occupants, the operating scenario and the on-site measurement data defined [72]. |
| | Smart City | Given the increasing availability of building data at the city level, digital twins can be a promising platform for building portfolio performance assessment and urban energy management [73]. The city can develop a digital twin to identify and simulate energy losses in buildings or improve building retrofits with a dynamic digital twin [74]. The resulting DT model can effectively categorize buildings based on usage, extract parameters that identify buildings within specific areas, and divide the city into energy consumption-related blocks to enable situational awareness and decision support for managing city infrastructure [21]. |
| Cost reductions | Manufacturing | The digital twin enables cost reductions by simulating the dynamic behavior of the industrial plant and predicting potential problems in the smart manufacturing paradigm [75]. |
| | | Since the digital twin has the capability to test, experiment and inspect a product under different conditions and usage environments ,there is no need to always build expensive prototypes and models, resulting in significant cost savings in product development and testing [76]. |
| | Smart Ciy | The urban digital twin can help reduce costs by simulating what-if scenarios in the system and predicting future outcomes to assess how cities are likely to behave under different economic, environmental, and social conditions in order to take the most appropriate actions and identify causes of potential disruption [77]. |
| | Built Environment | Digital twins may help reduce costs by enabling virtual construction sequencing and logistics scenarios to familiarize workers with required tasks and reduce costly rework, and by using AI to predict maintenance activities and events to avoid unforeseen costs [78]. |
| | Healthcare | The digital twin can contribute to savings in time and costs by testing different treatment methods on the patient's digital twin to avoid costly treatments that do not respond positively to the patient [66]. |
| Increased performance | Built Environment | Features such as bidirectional data exchange and real-time self-management (e.g., self-awareness or self-optimization) distinguish the digital twin from other information modeling systems and enable performance gains, as the digital twin is able to "learn" and propose and simulate new scenarios before building an object, manufacturing and provisioning tools and heavy equipment, and performing the construction process [63]. |
| | Manufacturing | The digital twin can perceive and simulate the production process based on real-time data, while it can simultaneously understand, predict and optimize production performance based on historical data [79]. |
| | Smart City | The urban digital twin serves as a "self-recognizing, self-determining, self-organizing, self-executing and adaptive platform for city operations and maintenance" to aid cities in real-time remote monitoring and enable more effective decision-making. As a result, decision makers can conduct city management in a more orderly manner and citizens can participate in city management processes and monitor government decisions [80]. |
| | Supply Chain | The full coverage of all data, states, relationships, and behaviors of the logistics system and the distinctive analytical capabilities of the digital twin enable the use of diagnostic, predictive, and prescriptive methods with the goal of holistically improving logistics performance along the entire supply chain [58]. |
| Enhanced collaboration | Built Environment | Important information from all actors (e.g. design documentation, inventories, material specifications, and schedules) can be stored and processed with the digital twin to support decision making and improve collaboration [78]. |
| | Smart City | The urban digital twin has the potential to "tackle urban complexity by visualizing complex processes and dependencies in urban systems, simulating possible outcomes, and impacts and taking into account the heterogeneous needs and requirements of its citizens by enabling participatory and collaborative planning" [42]. |
| | Supply Chain | The digital twin facilitates supply chain coordination between project participants by predicting potential logistical risks and an accurate estimated time of arrival based on reliable simulation, enabling just-in-time module delivery [71]. |

transportation infrastructure [10]. Maritime supply chains are commonly referred to as "the connected series of activities pertaining to shipping services which is concerned with planning, coordinating and controlling freight cargoes from the point of origin to the point of destination" [87].

Table 4 identifies the role and characteristics of the port in the smart city and as part of the transport chain in supply chains by comparing two common definitions. It points out that the port is a central part of both the supply chain and the smart city, and therefore a possible DT solution should consist of elements that fulfill both city and supply chain requirements. Conversely, this also means that the port can draw on the experience and expertise of existing DT solutions and research in these two related areas when implementing DTs. The fusion of the port and the city in providing infrastructure critical to the port can be recognised in the port of Hamburg's smartBRIDGE project, where the city and the port are working together on a DT solution to maintain the most important bridge for the port serving most of its hinterland transport [46]. The provision of effective hinterland infrastructure is also crucial for the transport chain, as it is often a major bottleneck and responsible for up to 60% of the global maritime supply costs [89].

The complex interplay between the port, the (smart) city and the supply chain is also illustrated in the three layer model of freight transport by Wendel et al. in [90] and the modified version for the city context by Behrends in [91]. The main components from top to bottom are the freight flow (representing the supply chains consisting of nodes and links), the transport network (which converts transport demand into traffic) and the infrastructure as well as their interaction in the overall system. In this respect, the model serves to describe these three interrelated systems, which are interconnected in the transport market (matching of freight flow demand with transport flow supply) and in the traffic market (matching of transport flow demand with infrastructure supply) resulting in a multi-actor challenge [92]. An adapted version of the three layer model with respect to the port DT is presented in Figure 3. The DT of the port interacts directly and indirectly with all three layers, as it represents a node in the transport chain serving various supply chains, while simultaneously offering various transport services in the transport network (e.g. intermodal transfer and temporary storage), and extensively using urban infrastructure for the hinterland traffic. As such, a critical view of the port's DT from the perspective of the supply chain and the (smart) city is inevitable. In the following, we will therefore take a closer look at existing DT research approaches and solutions within the smart city and supply chain domains.

### A. THE SMART CITY DIGITAL TWIN

A city can be considered a "smart city" if "investments in human and social capital and traditional (transport) and modern (ICT) communication infrastructure fuel sustainable economic growth and a high quality of life, with a wise management of natural resources, through participatory gover-nance" [93]. The port directly affects the urban infrastructure as a (global) mobility provider, while also using the urban infrastructure for its port hinterland processes. According to Marsal-Llacuna et al. [94], the purpose of the smart city is "to improve urban performance by using data, information and information technologies (IT) to provide more efficient services to citizens, to monitor and optimize existing infrastructure, to increase collaboration among different economic actors, and to encourage innovative business models in both the private and public sectors." Consequently, a large body of work sees smart cities as an inherently technological approach based on effective data collection, analysis and communication to solve urban problems that address economic, social, environmental and governance issues. However problematic (see eg [95] and the references therein), this view is, without doubt, a dominating legacy. A DT can provide the necessary processing of data collection systems to formalise the physical twin (here the city) and provide analytics for governance, management and maintenance [96]. Transportation is often of particular importance to the smart city, as transportation is a critical cornerstone of a city's functionality, development, and prosperity. Their close linkage is evidenced by theories such as urban land rent theory and location theory, which conceptualize the relationship between transportation and urban land use [97]. Therefore, transportation has traditionally been the primary focus of applied urban modeling, and urban transportation remains an important focus today because of the domino effect of transportation-related challenges such as addressing congestion and improving air quality on electricity infrastructure due to the prevailing strategy of decarbonising transportation to improve air quality [98]. Ports play a central role in cities as part of their infrastructure for mainly freight transport but also sometimes passenger transport and ports generate hinterland traffic that intermingles with urban traffic, which requires intelligent planning of hinterland traffic [99]. Ports thereby directly influence the three parameters smart energy, smart mobility, and smart infrastructure. It is therefore not surprising that Hall [88] has already chosen the port as a central component of the smart city infrastructure. Many smart cities are therefore trying to integrate the port as an integral part of the smart city [100]. This is especially true for port cities such as Hamburg, Rotterdam, Amsterdam, Singapore, Genoa, Los Angeles, Stockholm, Hainan, Valencia, Antwerp, Shenzhen and Montreal, which are striving to use digital technologies in port logistics to help policy-makers, urban planners and administrators efficiently manage the flows of resources and information circulating in port cities and reduce the waste of unused resources [101]. To counteract this, the Hamburg Port Authority has installed 300 sensors on roads and bridges as early as 2011 to monitor and control road traffic in the port. These provide the Port Road Management Center with information about the status of bridges (e.g. open/closed) and traffic throughout the port. Based on this data, the traffic management system can make efficient decisions on the routing of traffic flows to optimise routes and thus avoid congestion in overlapping city-port areas [102].

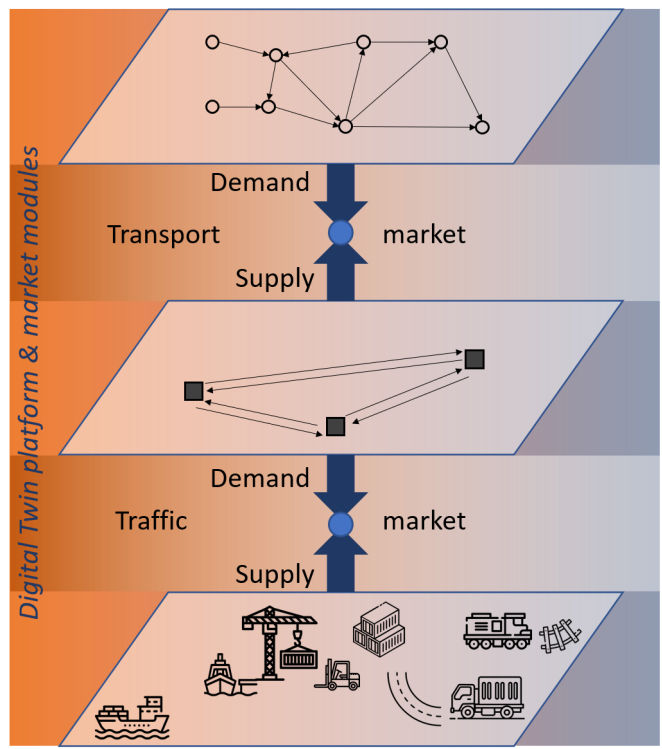

FIGURE 3: Digital Twin platform and market modules (adapted from [90])

TABLE 4: The two viewpoints of the port

| Port as | | Role of the port | Focus on |
|---|---|---|---|
| a supply chain node | "A seaport is a logistic and industrial node in global supply chains with a strong maritime character and a functional and spatial clustering of activities directly or indirectly linked to transportation, transformation, and information processes within global supply chains" [1] | Provider of services to the supply chain | Transshipment, Unloading operations, Modal shift, Information Hub |
| a smart city component | "A city that monitors and integrates conditions of all of its critical infrastructures, including roads, bridges, tunnels, rails, subways, airports, seaports, communications, water, power, even major buildings, can better optimize its resources, plan its preventive maintenance activities, and monitor security aspects while maximizing services to its citizens" [88] | Peri-urban infrastructure for freight and passenger transport | Mobility, Economic Activity |

The DT concept fits well with the notion of smart cities as complex systems that exhibit real-time behavior as well as emerge over the long term. The characteristics of DT seem to correspond to the need to control small contexts and environments and to be able to understand and control the macro behavior of the whole city [76]. It is therefore not surprising that the DT is seen as the ultimate tool for municipalities to design, plan and manage their networked infrastructures and assets efficiently and cost-effectively. In this context, it is predicted that the application of DT technology will enable savings of $280 billion in the design, construction and operation of cities by 2030 [103].

According to [104], the DT city pursues three visions: 1) more intensive and efficient urban production and operation; 2) liveable and comfortable urban living spaces; and 3) a sustainable urban ecological environment. Table 5 presents these three visions and provides guidelines and examples on how to realize them. A fundamental difference of the city DT compared to the conventional manufacturing DT is the magnitude of its scale. [21] states that the smart city is a "system of systems" and that the "the complexity and heterogeneity of city-scale DTs can be orders of magnitude greater than their industrial counterpart".

Similar to the port, energy is among the most important resources for operating a city; from large metropolitan areas to small towns, enormous amounts of energy are required for buildings, transportation, infrastructure, and services [21].

In summary, the city DT is intended to connect with the real city to improve the visibility of the city and the understanding and analysis of events and processes in the city. To this end, the city's DT is considered as an enabling technology to promote situational awareness for municipalities and to provide an information model for the city, i.e. the city's DT can collect, monitor and manage urban data. The city's DT is expected to accurately reflect and influence the city's functions and processes to improve its realisation, functioning and management [56]. At its best, the city DT contributes to a smarter and more sustainable city and enables citizens to enjoy a higher quality of life.

### B. THE SUPPLY CHAIN TWIN

Supply chains are commonly defined as "a network of connected and interdependent organisations mutually and cooperatively working together to control, manage and improve the flow of materials and information from suppliers to end users" [105]. In the context of supply chains, ports are often described as nodes in the transport chain of supply chains that connect organizational, logistics, and information networks to facilitate the seamless flow of goods [106]. The coordination of transport chains requires the synchronization and coupling of numerous consecutive transport and transshipment processes carried out by a multitude of actors such as freight forwarders, terminal operators, port authorities, shipping companies and railroad companies [60]. Ports and container terminals enable ports to add value to the cargo that passes through them, for example, ports can contribute to sourcing and pre-assembly activities [107].

TABLE 5: Three visions for the digital twin city, derived from [104]

| Smart City objectives | Urban DT contributions | How it is achieved | Example |
|---|---|---|---|
| More intensive and efficient urban production and operation | Digital twin technology is used to make an intelligent analysis of complex scenarios in terms of the flow of people, goods, energy and information. | Optimization of urban spatial layout, relieving traffic congestion at complex intersections, simulating and rehearsing responses to natural disasters, and scientifically formulating emergency evacuation plans | Virtual Rennes utilises 3D modelling and simulation techniques to optimise the decision-making process, helping urban planning departments to transform experience-based decisions into data-based decisions. |
| Liveable and convenient urban living spaces | Digital twin technology is used to monitor the performance of urban components, predict failures and avoid risks, and ensure the safety of residents. | In the digital twin city, residents can visualize urban congestion in real time to adjust their travel plans. Digital twin technologies can minimise the differences between online and offline experiences by creating hospitals, classrooms and community services through virtual-real interaction and customisation. | Georgetown's Digital Twin City platform assesses the feasibility of new infrastructure projects by simulating and extrapolating the potential impact and risks on existing infrastructure to improve urban liveability and to support land use reallocation and attract investment. |
| A sustainable urban ecological environment | Digital twin cities can help city managers and experts to: 1) evaluate and optimize ecological features; 2) make a comprehensive diagnosis of various carbon emission implementation policies and select optimal solutions; and 3) promote the efficient operation and maintenance of energy facilities and carbon trajectory tracking. | In the digital twin city, city managers can learn about the heat island effect, environmental pollution, climate change, energy use and other real-time conditions dynamically, analyse the gap in resource allocation among different regions, and automatically make policies that could optimize the flow and matching of resources. | The Virtual Singapore platform integrates data regarding ambient temperature and sunlight exposure throughout the day to visualize the heat island effect in urban areas as well as the effects of any potential measures taken in response. |

Wang et al. propose in paper [61] the creation of a DT-driven supply chain (DTSC) as an innovative and integrated solution for the smart supply chain (SSC), as the DT meets all the requirements of a smart supply chain. The selected criteria for smart supply chains and the elaborated results on how the DT can fulfill them are shown in Table 6. Although DTs are not yet widely used in logistics and supply chains, many of the key technologies are already in place. A number of relevant techniques are already used in logistics, including (1) a variety of sensors to track shipments and material handling equipment, (2) open API strategies and the shift to cloud-based IT systems, (3) machine learning and advanced analytics techniques for optimising supply chains and gaining new insights from historical shipping and operational data, and (4) implementation of augmented, mixed, and virtual reality applications [43]. Recent crises, such as the coronavirus outbreak with recurring COVID-19 lockdowns in China and the Russia-Ukraine conflict with wider geopolitical implications underlined the need to protect the supply chain from significant disruptions. In order to be prepared, firms must continuously monitor the the current course of events, check inventory and logistical hubs, and respond rapidly to changing circumstances. For a complex supply chain such as the maritime chain with a complex hinterland transportation structure, a DSCT can have many advantages. The DSCT could, among other things, (1) predict delivery times using machine learning from previous transports of goods through the supply chain, (2) identify the limits and bottlenecks of the supply chain by simulating different scenarios and workloads of the existing transport network, and (3) determine the most cost-effective measures to save energy or money through better collaboration and by making the best use of existing resources [60].

The first DT supply chain platform "Logistics Mirror" was developed by JD Logistics to increase supply chain efficiency and reduce costs. First deployed in JD's Beijing Asia No.1 logistics park, the platform now covers over 8,000 transportation routes across JD's own logistics and warehousing network, providing accurate capacity forecasts and alerts for sales peaks [108].

## V. PORT DIGITAL TWIN

The potential of DT´s to improve port processes towards energy savings and thus reducing costs or avoiding $CO_2$ emissions has been recognised by leading ports [11], [109], [110]. It is therefore not surprising that there is already a multitude of existing twinning solutions catering to the needs of different port stakeholders, twinning different port elements. For example, Zhou et al. in [111], discuss the implementation of a DT port crane framework based on multi-sensor data, which is able to reproduce the historical crane operation process, simulate the control program, simulate the synchronous mapping and take remote control. In [26], the authors present a decision support system with a DT-based resilience analysis that assesses a port's resilience to potential disruptive events, taking into account its design, operations and possible pre-defined post-event recovery measures to mitigate the impact of the disruption. Further decision support related DT studies emphasize on integrated crane maintenance under operation in container terminals [112] or on dispatching assistance in port logistics based on a performance forecast [113].

TABLE 6: DT contributions to achieve smart supply chain requirements, derived from [61]

| SSC objective | Explanation | Digital supply chain twin contributions |
|---|---|---|
| Connectivity | Connectivity is the ability to connect all enterprises, products, properties, and other valuable items in the supply chain in order to provide comprehensive information and to monitor the marketing status, intraenterprise operations, and inter-enterprise communications | In a DTSC, the physical supply chain is connected by smart sensors or online systems and collects specific data and information that enable the virtual supply chain to mirror the static properties and dynamic business processes of the physical supply chain. |
| Visibility | Visibility is the ability to keep track of the flow of materials, finances, and information throughout the supply chain. Supply chain managers must have access to real-time data related to production, inventory, logistics, and marketing in order to identify where and how products are stocked and when and how products are sold to consumers | Through real-time data collection and implementation, the DTSC is able to provide a link between the physical and virtual supply chains to synchronize operational dynamics, which increases supply chain visibility. |
| Agility | Agility is the quick ability to detect changes, collect relevant data, analyze opportunities and threats, make optimal decisions, implement these decisions, and modify operations accordingly | The synchronized data provide opportunities to monitor, analyze, control, and optimize the supply chain and results in up-to-date virtual simulation and optimization that provide agility |
| Integrated | An integrated supply chain shares information from all involved actors and manages all distribution and logistics activities through a central system, making decisions across different stages of the supply chain together | A DTSC actively operates in the entire business process throughout the supply chain and thus provides an integrated and holistic view that enables all stakeholders to collectively contribute to the creation across different stages of the supply chain |
| Intelligent | An intelligent supply chain makes largescale, optimal decisions and uses predictive analytics to protect the supply chain from future risks | A DTSC's predictive analytics, based on optimization and simulation algorithms, enable decision makers to look forward instead of backward and intelligently design the supply chain |

In [114], Li et al. propose a safety operation optimization framework integrating the DT with the AdaBoost algorithm [115] to increase container terminal efficiency and safety. In [116], Wang et al. present core techniques for a systematic framework of a DT-based model focusing on transport and operations for smart port management.

So far a large array of port DT research is limited to specific, isolated port processes or assets. However, a holistic port DT solution would need to take into consideration the port's relation with the smart city and supply chains, and thus consider the various transport flows along the transport chains of the numerous supply chains, as well as the hinterland processes through the urban infrastructure, not in isolation, but in their interconnectedness as a whole. Towards this, subsection V-A specifies the characteristics and requirements of a port DT by analyzing its core processes and procedures, based on interrelationships with the smart city and supply chain domains. Subsection V-B presents examples how energy savings can be identified through appropriate operating scenarios providing opportunities for optimal utilization of port facilities and equipment, and intelligent linking of port processes.

### A. THE PORT AS THE INTERSECTION OF SMART CITIES AND GLOBAL SUPPLY CHAINS

The aim of this section is to demonstrate how a holistic DT can significantly optimize the port as a whole, based on three core statements, which provide characteristics and requirements of a port DT which support the realization of the three DT "tools" of [12] focusing on situational awareness, intelligent decision making, and increasing the efficiency of collaboration between the numerous port stakeholders. These are discussed and demonstrated in detail on the basis of numerous application examples of the various port processes. In this context, suitable application examples and transferable insights from the DT application domains associated with the port, smart city and supply chain, are also discussed and applied to the port context. Therefore, the three statements below elaborate the necessary characteristics and requirements for a port DT that is not aimed at optimizing specific port assets or processes, but the port as a whole.

1) Through a high degree of data integrity and a virtual mirroring of all port processes, the DT enables a distinct situational awareness

Ports perform a variety of functions, being both nodes in transportation chains and locations of economic activities related to the handling of ships and cargo in the port. In addition to their main functions of (un)loading ships, connecting different modes of transport, and storing goods, many ports around the world have developed as locations for industrial activities and/or logistical operations, encompassing a large number of different, sometimes interrelated, industrial functions [7]. In addition to the Port's primary role as a global hub with the goal of establishing excellent port operations that enable the seamless transfer of goods between the maritime and hinterland networks, the Port also serves as an industrial cluster as well as an information hub [106]. Supply chain coordination, i.e., "the concerted action of the various actors in the supply chain to achieve better performance" [106], is driven by information sharing, leading ports to adopt information technologies, standards, and information systems to facilitate efficient planning, information sharing, and management of port activities and procedures [117].

Ports coordination task is complex with multiple stakeholders and processes and it is increasingly important to improve the overall view of these and identify potential bottlenecks to increase efficiency, safety, and sustainability [118]. Similar to the key objective of enhanced coordination of the city DT and the DTSC, the port DT should also have the capability to facilitate information sharing between all actors involved in the supply chains with transports passing the port as well as the port actors. The exchange of information (such as spatio-temporal status updates of ships approaching the port, operability of the port equipment, warehouse capacity, and degree of congestion of the inland transport system to the nearest inland terminal) leads to a higher situational awareness for all involved actors, while the analytic capacities of the DT can reduce the risk of bottlenecks and unseen events. Similar to the city DT, the port's DT should therefore also contain an interface that enables the seamless connection and interaction of all port actors and their respective operational processes. Therefore, the first fundamental step in the implementation of the port's DT is to create an interface that unifies the data of all port operators in order to achieve a collective inter-company awareness of the port. Besides a virtual presentation of the static port objects (quays), this also enables a visualisation of the dynamic processes, which consist of a multitude of different actors. Ideally, the synchronisation of all port processes in the DT will enable a virtual representation of all processes in almost real time, which will promote situational awareness. Depending on the nature of the process, updating is not always required in real time, many updates can be done at predefined intervals (e.g. environmental sensors provide regular minute-by-minute updates), others when certain events occur (e.g. the warehouse only gives a status update when a container has been moved or the terminal gate changes its status whenever a truck as entered or exit the terminal). The integration of both static and dynamic data, whether real-time or semi-real-time, can create additional situational awareness by providing knowledge about both less volatile base events (storage yard capacity or weather conditions) and high volatility events (e.g. energy consumption of critical infrastructure or the extend of traffic disruptions in hinterland traffic). Through standardised digital data exchange and subsequent dynamic visual representation through the DT, relevant port actors can gain the necessary situational awareness to make coordinated decisions that improve resource utilisation and enhance the safety and efficiency of operations to create added value for the individual port actor and the port as a whole. A DT implementation of the port should offer a high level of connectivity and visibility, similar to the supply chain DT, to best accommodate the transport flows that go through via the port (cross reference Table 6), while the level of detail and the dynamic nature of the ever-changing processes can be aligned with, for example, the traffic modules of the city's DT. The consecutive steps described here reflect the DT core aspects visualization and synchronization in Chapter 2.

### 2) The port DT should provide extensive data analytics capabilities for smart decision making

Terminal operators have a strong interest in terminal automation as it directly improves performance indicators such as cost, efficiency, safety and reliability [1]. Increased terminal automation is also needed to cope with the increasing size of freight ships and growing freight volumes. Therefore, different terminal processes are illustrated where the DT has potential to reach further levels of automatization. Each terminal is characterized by the interplay of sea, storage and transport operations. In the following, the respective processes that could be automated by implementing the DT for seaside operations, transportation operations, and yard operations, are briefly presented. Seaside operations are performed at the berth and the quay of the port. The three main decision problems considered at the seaside of the port are the Berth Allocation Problem (the allocation of berths to vessels), the Quay Crane Assignment Problem (the assignment of cranes to vessels for unloading and loading), and the Quay Crane Scheduling Problem that attempts to optimize the allocation and sequence of task distribution (unloading and loading) among a given number of cranes [119]. The most error-prone piece of equipment when operating at seaside is the quay crane. This is due, among other things, to the complexity of the tasks, the high workload of the cranes, which are often limited as a result of their high investment and maintenance costs, and the increasing precautions taken with regard to safety concerns and interference between quay cranes operating in parallel. Consequently, more recent quay crane scheduling approaches are increasingly integrating predictive maintenance to prevent breakdowns and save costs [120]. Since the DT can access both historical and real-time data, it offers ideal conditions for the integration of preventive maintenance into all operational processes. The goal of storage space allocation is to reduce the cycle time of storage yard operations (i.e., the time to store, retrieve, and reshuffle). The proficiency of a storage space allocation depends on the availability and quality of information about the arrival and departure times for the import, export and transshipment containers. Given the operating costs of gantry cranes and the imperfect nature of the information, containers are usually only reshuffled when needed [121]. Other common yard storage related problems are the dispatching (which equipment will perform which tasks) and routing (the sequence of tasks performed by each peace of equipment) of material handling equipment. There are various decision-making problems to be overcome when performing transport operations at the port. In paper [122], Iris et al. specify five decision problems and classify them into strategic and operational problems. At the strategic level, a terminal must decide on the type of vehicles to be used. In addition, decisions must be made about the number of vehicles to be used to perform all transportation operations related to each arriving vessel. At the operational level, routing and dispatch decisions must be made to associate tasks with vehicles and to provide transportation over specific routes from origin to destination. Avoiding collisions, congestion, and impasses are key attributes in routing decisions. The transportation process is usually designed in such a way that operations on the sea and land sides are coordinated to avoid bottlenecks [123]. It should be noted that the processes of these three operational sub-areas are closely interlinked in reality. For example, when the ship is unloaded, transfer vehicles must already be planned to transport the containers to the storage yard, which in turn ties in with storage yard planning.

A growing number of DT studies illustrate the potential of DT implementation to increase the effectiveness of certain port operations. These studies cover a wide range of different terminal operations and problems, including DT applications in automated storage yard scheduling [124], operating status monitoring for port cranes [111], and assisting truck dispatchers [113]. For example, the experimental results of a machine learning based DT implementation of the largest container terminal in Shanghai by Li et al. in paper [114] illustrate the potential of DT-based terminal operation. Through higher degree of automation and intelligent decision making, DT-based terminal operation achieves higher loading and unloading efficiency than conventional terminal operation, by 23.34% and 31.46% for small and large problems, respectively. This study demonstrates the DT capabilities of enhanced real-time optimization for increased operational efficiency and safety.

However, a holistic port digital twinning approach should include terminal operations across all areas (e.g., berth, quay, transport, yard, and gate) and assess how they are interlinked before taking decisions for individual components. Carlo et al. emphasize in paper [121] that synchronizing quay, transfer, and yard operations is of great importance to reduce vessels' turnaround times, which in turn leads to higher port performance.

### 3) The port DT as a facilitator of multi-stakeholder governance and collaboration

Similar to the smart city, the port is a hub of numerous processes involving multiple actors and dimensions. Ports not only align the interests of employees, management and shareholders, but also serve with a wide range of stakeholders, including terminal operators, vessel operators, railways shippers, industry associations, municipalities, and government agencies [125]. In the following, three port processes are presented that illustrate the complex interplay of the port's numerous actors that require a high degree of cooperation. The port call is one of the main processes of the port, which involves a number of actors and is necessary for the arrival of every ship. Upon entering the port, the port authority has to give its permission; pilots and tugs and other supporting nautical services are often needed to bring the ship from the port area to the berth; moorers tie up the ship at the berth; terminal operators and stevedores are involved with loading and unloading, while shipping agents make sure that everything goes according to plan [126]. Once the ship docks at the berth, proactive planning of unloading and reloading is required, not in isolation but in conjunction with other port processes. The process of (un)loading requires data about the ship and the cargo, intelligent management of the quay cranes, and scheduling of personnel for monitoring and onward transport within the port to the storage yard. For this purpose, an optimal use of the equipment and facilities in the port is needed. As the number of berth segments and quay cranes (QCs) in a port is generally very limited, proper coordination between berth allocation and QC allocation can improve the overall performance of the port. Following the two processes described above, including berth allocation and subsequent un(loading) operations, joint planning of yard processes is another essential sequential process. Yard management also relies on the interaction of all port stakeholders. When planning yard allocation for import and export cargo, ship status and yard status must be taken into account. Consequently, yard planning for export, import and transshipment containers as well as yard planning for empty containers and relocation within the container yard must be aligned, which requires a high degree of coordination [127].

The coordination of hinterland transport is also of key strategic importance to the port, as the hinterland offers the opportunity to add value-adding activities to the logistics supply chain, but also to strengthen the port's throughput by providing efficient inland transport to inland distribution centres [128]. In paper [129], van der Horst and de Langen emphasize the importance of coordination in the hinterland transport, because most bottlenecks of the door-to-door container transport chain, such as congestion, insufficient infrastructure, and problems with handling of barges, trains, and trucks at deep-sea terminals, occur in the hinterland network. In a subsequent work [130], van der Horst and de Langen conclude that the quality of hinterland connections is the result of joint actions by a large number of actors with a great deal of operational interdependence. These examples demonstrate that ports can learn and benefit greatly from the developments in the cities DT. According to Shahat et al. [56], a city DT depicts the interdependencies between its elements (such as people, infrastructure, and technologies) and the relationships between its domains on a platform, and provides a collaborative environment for the joint design and development of the city. They further argue that the integration of the city DT is done through two dimensions. The first dimension is the 3D model of the city itself and the second dimension is the provision of a collaborative environment for the various actors of the city. One example of the implementation of such a city DT is the Fishermans Bend DT for the Australian state government of Victoria [131]. The Fishermans Bend DT's primary purpose is to collectively improve public services, including road infrastructure and safety. Its simulation capabilities for real-time mass transit data facilitate the prediction of tram traffic patterns. The potential of the city DT for a higher degree of cooperation has also been highlighted in the context of Zurich and Herrenberg. Based on the expansion of the geodata infrastructure and the technical platform, the Zurich DT can be used for location-based cooperation with internal and external partners. This creates optimal conditions for the joint presentation, discussion and design of the public space, taking into account various analyses and calculations such as visibility, sound propagation and solar potential analyses, shading calculations and flooding simulations [44]. The Herrenberg DT aims to visualize the complex "invisible", such as urban data and simulations, to support citizen participation or expert collaboration [42].

The principle of two dimensions to support collective decision making in the port can also be applied to the port. The first dimension would thus be a 3D representation of all port buildings and a dynamic illustration of all port processes as described in chapter V-A1, while the second dimension would provide a platform for collaborative decision making, in which all port actors are networked, and simulations of what if scenarios predict the effects of different measures on the actor itself, but also on the port as a whole. In addition, the various port stakeholders could already report known difficulties, such as staff shortages or planned maintenance in the platform, whereupon the DT's simulation models would already indicate possible consequences and alert the other stakeholders to the potential impact. In the case of port expansion or the introduction of a potential new policy, the DT platform would also provide all stakeholders with insight into the consequences, thus supporting joint decision-making. Therefore, the various stakeholders could be actively involved in the decision-making process through suggestions and direct testing of these.

In the case of the hinterland aspect, a port's DT could recommend to port decision makers the best initiatives to increase hinterland transport efficiency in order to reduce urban congestion and CO2 emissions. Thus, known countermeasures such as terminal systems, extended gate opening hours and virtual container depot systems could be automatically regulated [132].

## B. EXAMPLES OF EFFICIENCY GAINS WITH PORT DIGITAL TWINS

Ports consume a lot of energy as they play a central role in the world's physical infrastructure, handling 90 % of the world's cargo flows. This is particularly evident in the case of the port of Hamburg, which accounts for around 40 % of the city's total energy consumption [133]. Port operations are thus very energy-intensive activities, putting port authorities and terminal operators under great pressure to meet economic and environmental standards and to make container terminals not only more competitive and productive, but also more sustainable [134]. Together with air quality and climate change, energy efficiency is one of the top 3 environmental priorities of the European Sea Ports Organisation of which all three are interrelated to effectively promote sustainability [135]. As nodes in extensive global transport networks and as intersections of large supply chains, ports cause environmental impacts and negative effects on the Earth's climate through their logistical function (i.e., transportation, terminal, and warehousing) and their industrial and semi-industrial function (i.e., goods and energy production, assembly, and recycling) [136]. A port's operational efficiency depends on how efficiently the available resources are managed. Therefore, there is a positive correlation between the reduction of operating times (e.g. ship handling times, transport times of containers in the yard) and operational efficiency in ports [137]. The DT could apply optimization and machine learning to optimize the efficiency of the various port processes and port facilities to save time and energy. This would result in reduced turn around times, fewer queues at the port, and more energy-optimized ship movements before entering the port. A comprehensive overview of port processes and activities whose operational optimization could lead to energy savings using the DT is depicted in Table 7.

In the following, some steps are presented on how a DT implementation can aid the port to increase its operational performance while reducing C02 emissions. First, based on the principles of the Internet of Things, a ubiquitous sensor network can collect real-time data on port traffic (sea and land), the ecological environment, and the various processes of port operations, enabling a connection and mapping from the physical world to the digital world. Second, based on the collected operational data of the port and its evaluation by the data analytics-driven models of the DT, decision makers can be assisted by simulating what if scenarios in the digital space to find the most efficient designs and set ups. Third, through ubiquitous data analysis, decisions could be made automatically and passed on to port facilities and equipment without human interaction, shortening decision times and minimizing the risk of human error. Ideally, two-way data integration and interaction could be achieved, allowing all decisions to be made remotely based on the DT simulations, thus embedding the bi-directional nature of the DT into the port. In the best case, the DT should be able to make decisions regarding operations and maintenance autonomously based on real-time and historical data and their further processing in its models. The DT should be able to integrate preferences of port decision makers at any time, allowing them to change different parameters for the respective configuration and thus achieve the best data driven result for different application objectives (e.g. energy savings, max throughput). All processes in the port should always be considered in their overall picture (e.g. how different port processes influence each other, what is the impact on the lifetime of the equipment used and so on). Consequently, a holistic DT would enable complete self-governance while providing total oversight and transparency. The consecutive steps described here reflect the DT core aspects modelling and synchronization in chapter III-B. The implementation of the port DT can capitalise upon the experience of existing DT solutions in the smart city context, which also aim to analyse and improve a wide range of complex and interrelated processes (cross referencing section IV-A). In addition, the port can simultaneously benefit considering some of the objectives of the digital supply chain twin (e.g. integration and intelligence in Table 6) and at the same time pay attention to supporting the supply chain in its objectives to increase competitiveness.

TABLE 7: Areas where digital twins can help increase efficiency

| Port area | Port processes | Digital Twin contributions | References |
|---|---|---|---|
| Vessels navigating in the port or at berth | The virtual arrival time manages the speed of the ships according to the existing and imminent situation at the berth of the port so that the ship arrives without anchoring. This approach ensures "just in time" management of traffic and a reduction in fuel consumption for the ships. | Connected ships in a port regulated by a digital twin can operate "autonomously" and communicate with each other to avoid collisions and reduce their speed when the port is congested to save $CO_2$ and avoid queues outside the port. | [138], [139] |
| Port operations | Port equipment, especially quay cranes, are large energy consumers. In addition to energy-efficient scheduling, further tangible savings can be achieved by operating several cranes simultaneously to balance their consumption and by synchronising their movements to further reduce consumption. | The digital twin could facilitate the automation of cargo handling equipment by collecting and processing operational and energy data in real time. In this way, the digital twin can reduce energy consumption in real time by intelligently linking the different devices and integrating them into the port processes to save energy. | [137], [140] |
| | Improved port infrastructure utilisation reduces port congestion and contributes to a better energy balance of the port. Optimal use of space and equipment, an efficient gate processing system and the use of extended gate operating hours are key factors. | Based on a comprehensive data collection of all current port operations and an estimation of future events, the digital twin can help to continuously make the best use of limited port resources and aid decision makers in implementing the best policies. | [141], [142] |
| | The port's ultimate goal is to minimise the berthing times of ships, the resources needed to handle the workload, the waiting times of customer trucks and the congestion on the roads and in the storage blocks and docks within the terminal, and to make optimum use of the storage space. The latter is particularly dependent on proactive storage of containers in the yard terminal. | Through the holistic integration of all foreland, hinterland and port processes, the digital twin can contribute to a predictive stacking and storage of containers, with the aim of avoiding restacking and congestion in the port. Optimised yard scheduling by the digital twin can effectively reduce waiting times and limit the waste of resources. | [124], [143] |
| | Preventive maintenance and its implementation during periods of low berth utilisation minimises the risk of breakdowns, enables constant energy-efficient operation and reduces the risk of energy-intensive interventions in port operations. | The digital twin can optimize the trade-off between maintenance and maximum utilisation of port equipment by intelligently evaluating data in real time and comparing it with past events and patterns. | [112], [144] |
| Hinterland transport | Improving the efficiency of truck operations in the port, e.g. by optimising truck arrivals through the use of truck appointment systems, can contribute to improved energy efficiency while reducing port emissions. | By mirroring all port operations, the digital twin enables increased situational awareness in the port, allowing the arrival of trucks and trains to be planned in advance to avoid congestion. | [145], [146] |
| | The optimisation of the routing of vehicles can reduce energy directly through short routes and indirectly through the avoidance of congestion. | The DT-based models can guide trucks and trains in the ports in an energy-saving and congestion-avoiding way. | [141], [147] |
| Port facilities | In large port areas, lighting consumes a lot of energy and incurs high costs, as many ports operate around the clock and require adequate lighting. Optimising lighting systems (and the corresponding energy consumption) based on real-time data enables savings in both energy and costs. | The DT-controlled management dynamically monitors the port's lighting system and ensures that the lights are only switched on when vehicles approach the docks. Valencia Port Authority saves nearly 80% of energy consumption after implementing dynamic lighting system | [116], [117] |
| | Measures for real-time monitoring of energy consumption of port buildings and other equipment (e.g. reefer containers) to support decision-making and implementation of energy efficiency measures are not only relevant for terminals, but also for all industrial and logistic sites in the port area. | With the aid of sensors installed in the buildings and port infrastructures, the digital twin model can progressively collect data and provide a virtual replica of reality at any time. It is then possible to identify areas with high energy loads or those with the highest priority for energy renovations. In addition, the energy supply can be automatically reduced for buildings with low utilisation. | [148], [149] |

## VI. PORT DIGITAL TWIN IMPLEMENTATION CHALLENGES

Despite the steady growth of DT implementations across various domains, there are still a number of challenges involved in creating and integrating DTs into complex systems such as the port. These difficulties include technical aspects, organizational barriers, financial bottlenecks, and the risk of a further divergence between technologically strong and weak ports. The following list provides an overview of the main challenges.

- **Greater discrepancy between ports:** An increasing level of digitization of small and medium-sized ports can help improve their competitive position in the market for maritime transport services and strengthen the operation of the various sea-shore transport chains. However, ports with lower annual cargo throughput have limited opportunities to digitize port operations and related activities. This fact may be caused by limited financial and organizational capabilities to make large-scale investments [150]. It can therefore be assumed that the increasing development of DTs will further increase the discrepancy between financial and technical strong and weak ports.
- **Affordability:** Another barrier to implementing a port DT is the huge initial investment required for the underlying technologies, including sensor technology, information technology infrastructure, equipment suitable for automation, and the creation of the platform for the DT itself [151]. In addition, new skilled staff will also be needed.
- **Insufficient data capacities:** A DT's accuracy and precision of representation for an object, which forms the basis for decision support, depends on the available, accessible and compatible data, as well as the complexity of the object to be represented. Consequently, it can be concluded that the representations and predictions of a DT are only as good as its underlying design and operating data allow it to be [131].
- **Complexity constraints:** Complex port processes with numerous parameters lead to a complex multi-objective optimization problem [152]. Depending on the size and interconnectedness of the problem, e.g., joint optimization of a berth allocation and quay crane assignment under consideration of maintenance activities [120], the number of parameters included in the machine learning optimization problem can be enormous.
- **Security and data privacy:** Given the use of IoT and cloud computing, special attention must be paid to robustness against hacking and viruses when developing a DT environment [81]. Hacking private, confidential, or valuable information could harm all sources and parties involved in the complex network of port actors.

- **Unwillingness to share data and data sovereignty:** Despite potentially competing business goals, cooperation and collaboration are required to share and create data and models in a multi-stakeholder environment such as the port, where numerous processes are interconnected. Fairness, data security, and intellectual property rights must be ensured among stakeholders [153]. This relates in particular to concerns about insufficient data sovereignty. To ensure self-determination over data, there is growing interest in international data spaces. These describe a software architecture for enforcing data sovereignty in enterprise ecosystems and value networks [154].
- **Standardization:** To realize the full potential of DTs, especially for applications such as the port, whose complex interplay of dynamic processes and actors highlights the need to standardize a wide range of technologies, from data collection to insight extraction to decision making. The process of developing standards tends to be slow, leading to slow adoption of DTs on a broad scale [153].
- **Changes during operation:** To accommodate the dynamic evolution of systems and their environment, DTs must be designed to add new models or data as they become available. The challenge here is to develop architectures and frameworks for DTs that support such an open environment and provide functions for integrating new models and data while the system is in operation [155]. This is particularly important for the port, since the port is characterized by a constant inflow and outflow of ships in the foreland and vehicles in the hinterland, which must be integrated with or removed from the DT.

## VII. DISCUSSION

In a recent paper, Min [151] concludes that "the ultimate goal of a smart port is to make the port truly intelligent and enhance its self-learning capabilities". The analysis of the literature review presented in this paper reveals that the port DT can significantly contribute towards the fulfillment of this goal, especially the DT characteristics listed in III-B, as well as the discussion of the requirements and specifications for the port DT in section V. By continuously capturing real-time data and extracting patterns and insights from historical data, the port DT can identify operational patterns in real-time and historic data which can help identify inefficiencies and future breakdowns and thus enable taking preventive countermeasures to avoid them. This can be both at the process level, where the DT can recognize when maintenance is due to prevent bottlenecks, but also at the system level, where the DT, based on the evaluation of all processes within the system can help to select the best time for maintenance in order to minimize the loss of performance due to downtimes. Ideally, the port DT should also be able to initiate corrective actions related to operational scenarios in the event of cancellations or delays of arriving vessels or inland trucks. In addition, the port DT can be expected to support future technological developments of the port and transportation in general, such as the guidance of autonomous vehicles, port-wide human-robot collaboration and virtually guided self-service [151].

As demonstrated in section IV, ports are closely linked to the various transport chains of supply chains that use port services on their way from origin to destination [1]. A similar level of connectivity exists with cities, as the port has a significant economic and environmental impact on them as a mobility provider and driver of port-related industries that boost urban economies, while also leveraging their infrastructure to move cargo through the port hinterland to its final inland destination [8]. As highlighted throughout this article, this close functional linkage, as well as the fact that ports, cities, and supply chains are all characterized by a complex network of interconnected processes involving numerous actors, offers all three systems the opportunity to draw from each other's experience with DT solutions. At the same time, this insight leads to the question of how the corresponding DT solutions of all systems can be linked in the best possible way in order to increase the overall system of systems performance.

The assessment of early port DT contributions reveals that a large part of these initial DT solutions in ports portray DTs as a tool consisting of a set of systems that collect, process, and visualize data [45]. Although this view of DTs reflects some fundamental DT assumptions such as digital replication and physical-digital connections [34], such a perception of a DT cannot be clearly separated from a digital shadow [63] and therefore contains only certain parts of DT characteristics. Recent papers outlining the maturity of digital twins in the built environment [51], [78] and an adapted DT maturity assessment for ports in [52] reveal that mature DT solutions go far beyond real-time visualization, as they interact directly with their physical counterpart in the best case, leading to real-time optimization of specific assets and execution of the best overall runtime scenario determined by the DT models, while reducing costs and increasing safety through enhanced automation. With the increasing number of new DT solutions for ports and the continuous development of existing solutions, there is a growing opportunity for port DTs to gain adaptive awareness through the increasing accumulation of data and knowledge based on past decisions and thus learn from suboptimal decisions and their consequences and causes in the past.

The review of previous studies demonstrates that the potential efficiency gains of port DTs, especially in terminal operations, is acknowledged [113], [114], [156], and this will help to reduce costs [157], perform preventive maintenance [80], contribute towards port resilience [158], and increase operational safety [116]. However, the discussion of DTs in the port context in terms of energy savings through energy-aware scheduling of equipment, optimal use of port facilities, and due to operational efficiency gains through implementation of the best operating scenarios has been little explored in research. So far, mainly the potential for energy monitoring has been identified, together with the potential to save energy through dynamic lighting [116]. The conducted assessment of the potential of the port DT in terms of energy savings through energy efficient operational scenarios, presented in Table 7, indicates that DTs have the potential to contribute to fossil-free transport and therefore motivates the need for further research investigations in this regard in future studies. Compared to previous port studies regarding the potential of DTs, this study approached the port as a system, observing the port in its full functionality in terms of its interconnected processes, but also with respect to its interconnection with other complex systems, such as the supply chain, and the smart city. The DT characteristics derived in the course of this paper can be considered as a cross-domain guideline for DT characterization, but also, together with the presented specifications and requirements for the port, as a basic framework for the consideration of DT implementation. The port's DT provides comprehensive situational awareness, offers extensive data analytics capabilities for intelligent decision-making and provides an interface for governance and collaboration between different stakeholders, allowing the port's DT to make decisions autonomously but with full transparency. Section V demonstrated in detail that the port DT differs significantly from traditional applications of DTs, including manufacturing, in terms of its complexity and magnitude of interconnected processes, participating actors, and on-going changes.

## VIII. FUTURE WORKS
### A. EXPECTATIONS OF THIS WORK FOR THE FUTURE
This article aims to add value in both DTs' and the ports' future development. From the DT development perspective, the numerous port processes in section V-A and the port operational strategies to reduce energy consumption in section V-B provide the necessary understanding for DT requirements and use cases from the port's perspective. Furthermore, by comparing numerous DT definitions and purposes and assessing them for commonalities, the article contributes to a better understanding of DTs and their characteristics. From the port perspective, identifying the numerous use cases and operational DT-based strategies presented to obtain efficiency gains and reduce energy consumption may have a positive impact on port operators to recognize the value of DTs for their ports. The resulting port digitization efforts, in turn, accelerate the development of more sophisticated DT solutions. Thus, this article's importance lies in enabling both academics and practitioners to draw on the insights it provides from seeing the port jointly from the perspectives of smart city and supply chain, when considering the ports digital twinning process.

### B. FUTURE RESEARCH
Towards enabling more sophisticated and meaningful digital twin solutions, we discuss that further research is needed in two directions, relevant of our work:

### 1) Overcoming barriers in the creation of complex DT solutions, such as in the case of ports

The challenges presented in section VI indicate that there are several barriers to the adoption of port DTs. Moreover, previous research revealed that digital initiatives - even if innovative and targeted - especially in complex systems such as ports, can easily fail if the different requirements, perspectives and impacts on individual stakeholders are not adequately considered [159]. In addition, recent studies indicate that there is a general tendency in the transport industry to maintain traditional business practices and a reluctance to change conventional processes, resulting in a reluctance to pursue innovative services or process innovations [160]. Enabling research should therefore focus on overcoming the barriers presented in section VI. More precisely, we recommend further research within the following two directions:

- **Policies for overcoming barriers in organizational structures and processes:** The comprehensive review by Wohlleber et al. in [161] on port and logistics-related challenges to the adoption of DT solutions reveals that there are several organizational barriers that hinder the adoption of DTs on a larger scale. These challenges include lack of coordination and collaboration, heterogeneous organizational structures or cultures, high implementation costs and risks, lack of capabilities to change, lack of support from stakeholders, and lack of trust in DTs. Future research is therefore needed to understand the reasons for these barriers and to find solutions to overcome them.
- **Technological solutions to manage the multitude of input parameters and data in complex systems such as the port:** Semantic technologies such as ontologies have a high potential to realize the interconnection of all information and to ensure the openness of the DT approach to add further artifacts at any time [162]. This is particularly important for ports, as they are characterized by the dynamic coupling and decoupling of numerous artifacts, such as ships, trucks or other temporary resources. Another promising solution approach is the integration of DTs and digital threats, as this enables the integration of data into a platform that allows seamless use and easy access to all data [163]. The use of digital threats could enable the required involved port players to link their data through common inputs and data flows, facilitating integrated models and holistic data-driven decision-making. Thus, information sharing among normally isolated port companies would be enabled to facilitate a more time and cost efficient twinning process of the port as a whole.

### 2) Improving the complex interplay between the port DT and DTs of related domains, especially smart cities and supply chains

Since the port is also an important node in supply chains transport chains, as well as a mobility service provider and an economic hub of a smart city, future research should thus also address how the port DT should be linked to supply chain and city DTs and identify areas of interaction. Potential areas of interaction with smart cities may include (1) energy management as ports are large consumers of energy [164], (2) maintenance of shared infrastructure [165], and (3) measures to avoid congestion in the course of urban and hinterland transport [89], thus ensuring smooth urban traffic and efficient hinterland transport. As with the smart city, there are many synergies between the port and supply chains, including the rising need for the systematic use of greener alternatives for logistics, which affects the way ports operate and connect with each other, transforming supply chains in favor of modal shift and synchronicity [1]. Further research is therefore advisable to identify more precisely the intersections between port, supply chains and smart city, and to investigate their joint management in their respective DT systems, e.g. sharing of hinterland infrastructure, as in the case of the Köhlbrandbrücke in Hamburg [46].

As a result, there is a need for further research to create DT standards necessary for consistency in modeling and simulation of objects across industry silos and relevant domains [166]. This is compounded by the fact that due to the multitude of available different technologies at the digital twin level, and even different understandings of what it consists of, it is difficult to produce standards in a timely manner [167]. Consequently, future research should focus on harmonizing systems so that components of the same system and those of connected systems have, if not identical formats, at least formats that conform to harmonized rules [167]. Recent research therefore underscores the importance of interoperability, as its absence at the very least contributes to greater barriers, risks, inefficiencies, and even outright failures [168]. In this context, DT ontology is very important because, if chosen correctly, it enables effective DT interconnection and the creation of a twin network [169].

## IX. CONCLUSION

The paper dealt with the question of what constitutes a port DT. For this purpose, an extensive literature review was conducted on definitions and characteristics of DTs in various domains, taking into account their transferability to the port. More precisely, the paper specifies what constitutes a DT of the port, what requirements it has to meet, and presents concrete measures on how the DT can help to save energy through efficiency gains in use of port facilities and equipment, as well as intelligent process optimization. We therefore close the paper by proposing the following as a port DT: A digital twin of a port is a grouping of models and algorithmic components that jointly describe the complex interplay of port processes and operations allowing the characterization, estimation, and prediction of the most efficient operations at the process level, but also for the port as a whole. Through inputs from real-time sensors and experience from historical data, a user can identify patterns that led to inefficiencies in the past, get a complete view of current operating conditions, and predict future conditions by simulating what-if scenarios. Moreover, the algorithmic components of the port DT may allow it to act autonomously at any time, while providing full transparency, enabling the port to become a self-adapting system.

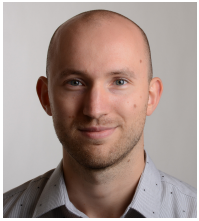

ROBERT KLAR received the M.Sc. degree in transportation and geoinformation technology from the Royal Institute of Technology (KTH), Stockholm, Sweden, in 2021. He is currently pursuing the Ph.D. degree at the Department of Science and Technology, Linköping University (LiU), Norrköping, Sweden, and is also employed as a Research Assistant at the Swedish National Road and Transport Research Institute (VTI), Linköping. From February 2021 to August 2021, he worked as a Research Assistant within the Division of Geoinformatics, KTH. His research interests include digital twins with respect to ports and the application of machine learning and optimization towards efficient and sustainable port operations.

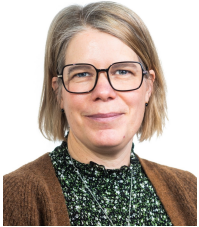

ANNA FREDRIKSSON received a Ph.D. degree from Chalmers University of Technology, Sweden and a docent (Habilitation) degree from Linköping University, where she is currently a Professor in Construction logistics and Head of Research Education at the Department of Science and Technology. She has a research interest in material flow management and production logistics within different industries and leads several projects focusing on decreasing the environmental impact of freight transport and improving efficiency of logistics in general. She is an associate editor of Journal of Manufacturing Technology Management.

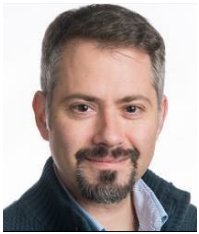

VANGELIS ANGELAKIS received a Ph.D. degree from the Department of Computer Science, University of Crete, Greece, in 2008. In 2015, he received a Habilitation (Docent) from the Department of Science and Technology (ITN), Linköping University (LiU), Norrköping, Sweden. He is currently a Professor with ITN, LiU. He has authored over 80 papers in international journals and peer-reviewed conferences. His research interests, departing from the field of communication systems and their application in the IoT, lie in the area of inclusive and sustainable smart cities solutions design, implementation, and evaluation.

He has held multiple guest posts both in the industry and academia, primarily through EU funded projects, in US, EU, and China. He coordinated and was a Principal Investigator for LiU in a host of EU projects since FP7, and has been with the OPTICWISE COST Action Management Committee. Currently he is LiU PI for two Horizon Europe projects. He served as an Associate Editor (AE) for the IEEE/KICS Journal of communications and Networks, and is serving as AE the Elsevier Internet of Things and the IET Smart Cities journals. He has been serving as an organizer and technical program committee member in a wide range of international conferences and workshops.

. . .